\documentclass[superscriptaddress,secnumarabic,twocolumn,amssymb,amsmath,nobibnotes,aps,prd,showkeys,nofootinbib]{revtex4}
\usepackage{}%
\usepackage{graphicx}
\usepackage{epsf}
\usepackage{bm}
\usepackage{amsmath}
\usepackage{amsfonts}
\usepackage{amssymb}
\usepackage{epstopdf}%
\usepackage{color}
\begin{document}

\title{Dark sectors with dynamical coupling}

\author{Weiqiang Yang}
\email{d11102004@163.com}
\affiliation{Department of Physics, Liaoning Normal University, Dalian, 116029, P. R. China}

\author{Olga Mena}
\email{omena@ific.uv.es }
\affiliation{IFIC, Universidad de Valencia-CSIC, 46071, Valencia, Spain}

\author{Supriya Pan}
\email{supriya.maths@presiuniv.ac.in}
\affiliation{Department of Mathematics, Presidency University, 86/1 College Street, Kolkata 700073, India}

\author{Eleonora Di Valentino}
\email{eleonora.divalentino@manchester.ac.uk}
\affiliation{Jodrell Bank Center for Astrophysics, School of Physics and Astronomy, University of Manchester, Oxford Road, Manchester, M13 9PL, UK}

\begin{abstract}
Coupled dark matter-dark energy scenarios are modeled via a dimensionless parameter $\xi$, which controls the strength of their interaction. While this coupling is commonly assumed to be constant, there is no underlying physical law or symmetry that forbids a time-dependent $\xi$ parameter. The most general and complete interacting scenarios between the two dark sectors should therefore allow for such a possibility, and it is the main purpose of this study to constrain two possible and well-motivated coupled cosmologies by means of the most recent and accurate early and late-time universe observations. We find that CMB data alone prefers $\xi(z) >0$ and therefore a smaller amount of dark matter, alleviating some crucial and well-known cosmological data tensions. An objective assessment of the Bayesian evidence for the coupled models explored here shows no particular preference for the presence of a dynamical dark sector coupling.

\end{abstract}

\maketitle

\section{Introduction}

Dark matter and dark energy, according to a series of observational evidences, are the two main constituents of the universe, comprising nearly 96\% of its total energy density~\cite{Aghanim:2018eyx}. For the last twenty years, a huge observational effort has been devoted to unravel the nature of these two fluids~\cite{Suzuki:2011hu,Crocce:2015xpb,Alam:2015mbd,Hinshaw:2012aka,Ade:2015xua}. Despite the fact that some of their properties have been measured with unprecedented accuracy (the value of the dark energy equation of state with $95\%$~CL errors is $w=-1.028^{+0.063}_{-0.061}$ from the latest Cosmic Microwave Background (CMB) data, combined with large scale structure observations and Supernovae Ia luminosity distances~\cite{Aghanim:2018eyx}), their nature still remains obscure. Furthermore, the so-called \emph{why now} problem provides another puzzle that may suggest a contemporary evolution of the two dark fluids. From the particle physics perspective, if a cosmic scalar field is responsible for the dark energy component, it may couple to all other fields in nature, if it is present~\cite{Carroll:1998zi}. These models emerged as coupled quintessence~\cite{Wetterich:1994bg,Amendola-ide1,Amendola-ide2,Pavon:2005yx,delCampo:2008sr,delCampo:2008jx}. Indeed, the presence of an interaction between the two dark fluids could successfully address the cosmic coincidence problem. Furthermore, some quintessence models could also be interpreted as modified gravity (Brans-Dicke-like) theories. An extra bonus supporting interactions among the two dark sectors arises from the fact that, when dark matter and dark energy interact, an effective equation of state $w < -1$  could naturally appear~\cite{Sadjadi:2006qb,Das:2005yj,Pan:2014afa}. While plenty of work in the literature has been devoted to explore the rich phenomenology of these  models~
\cite{Billyard:2000bh,Barrow:2006hia,Amendola:2006dg,He:2008tn,Valiviita:2008iv,Gavela:2009cy,Majerotto:2009np,Gavela:2010tm,Clemson:2011an,Pan:2012ki,Pan:2013rha,Yang:2014vza,Yang:2014gza,Pan:2016ngu,Mukherjee:2016shl,Sharov:2017iue,Yang:2017yme,Yang:2017ccc,Pan:2017ent,Yang:2018pej,Yang:2018ubt,Yang:2018xlt,Martinelli:2019dau,Paliathanasis:2019hbi,Pan:2019jqh,Yang:2019bpr,Yang:2019vni}, more recently, an extra encouraging aspect of these theories has improved their role as an alternative to a pure $\Lambda$CDM universe. Namely, in dark matter-dark energy coupled cosmologies the tension between local and CMB estimations of the Hubble constant $H_0$ could be ameliorated~\cite{Kumar:2017dnp,DiValentino:2017iww,Yang:2018euj,Yang:2018uae}.

Current cosmological observations still allow for significant interactions among the two dark sectors, i.e. between dark matter and dark energy, see e.g. Refs.~\cite{Salvatelli:2014zta, Nunes:2016dlj, Kumar:2016zpg, vandeBruck:2016hpz, Yang:2016evp,Yang:2017zjs}. In this work we consider an interacting scenario in which vacuum interacts with pressure-less dark matter adopting the more general phenomenological viewpoint, i.e. inspecting a time-dependent coupling. Such a consideration also entails the case of a coupling parameter that remains constant in cosmic time. For our analyses we have assumed that our  universe is homogeneous and isotropic, that is, its geometry is well described by the Friedmann-Lema\^{i}tre-Robertson-Walker line element.

The work has been organized as follows. Section~\ref{sec-efe} contains the gravitational equations within an interacting universe. In Sec.~\ref{sec-data} we describe the observational data and the methodology used to constrain the interacting dark energy models. Section~\ref{sec-results} presents the observational constraints on the models, including also a Bayesian evidence analysis. Finally, we draw our conclusions in Sec.~\ref{sec-conclu}.

\section{Gravitational equations of a universe with interacting dark sectors}
\label{sec-efe}

A homogeneous and isotropic universe is well described by the Friedmann-Lema\^{i}tre-Robertson-Walker (FLRW) metric:

\begin{eqnarray}
ds^2 =  - dt^2 + a^2 (t) \left[\frac{dr^2}{1-\kappa r^2} + r^2 \left(d\theta ^2 + \sin^2 \theta d \phi^2 \right) \right]~,  
\end{eqnarray}
in which $a (t)$ is the expansion scale factor of the universe and $\kappa = 0$, $+1$, and $-1$ correspond to a spatially flat, closed and open universe, respectively. In the following, we shall assume that the gravitational sector of the theory is described by the Einstein gravity and $\kappa=0$. Within this simple cosmological scenario, we will introduce  an interaction between the pressure-less cold dark matter component and the dark energy fluid, acting as vacuum energy. All in all, the conservation equations read as
\begin{eqnarray}
\nabla^\mu (T^{\rm CDM}_{\mu\nu} + T^{\rm DE}_{\mu\nu})= 0~,\label{conservation}
\end{eqnarray}
where $T^{i}_{\mu \nu}$ ($i = {\rm CDM}, {\rm DE}$) is the energy-momentum tensor for the $i$-th dark sector sector. Considering the dark energy fluid as a cosmological constant 
with an equation of state $w_x = p_x /\rho_x = -1$, the conservation equations are given by  
\begin{eqnarray}
&&\dot{\rho}_x = Q,\label{cons1}\\
&&\dot{\rho}_c = - 3 H \rho_c - Q,\label{cons2}
\end{eqnarray}
where the dot refers to derivatives respect to the time $t$, $H \equiv \dot{a}/a$ is the Hubble parameter, $\rho_c$ is the cold dark matter mass-energy density and $Q$ encodes the interaction rate between the dark fluids. Our analyses will be applied to two possible models, named as Interacting Vacuum Scenario 1 and 2 (IVS1 and IVS2, respectively):

\begin{eqnarray}
&\rm{IVS1}:& Q = 3 \xi (a) H \rho_{x}, \label{ivs1}\\
&\rm{IVS2}:& Q= 3 \xi (a) H \frac{\rho_c \rho_{x}}{\rho_c+ \rho_{x}},\label{ivs2}
\end{eqnarray}
where $\xi (a)$ is a time-dependent dimensionless coupling. 
A Taylor expansion of $\xi (a)$ around the present time ($a = 1$) leads to  
\begin{eqnarray}
\xi (a) = \xi_0 + (a-1) \xi^{\prime} (a_0 =1) + \frac{(a-1)^2}{2!} \xi^{\prime \prime} (a_0 =1) +...
\end{eqnarray}
where the prime stands for the derivative with respect to the scale factor. In this work, we shall restrict ourselves to linear corrections of $\xi (a)$. We therefore consider the following parameterization of the time dependent coupling parameter 
\begin{eqnarray}
\xi (a) = \xi_0 + \xi_a \; (1-a)~, 
\end{eqnarray}
where we use $\xi_a$  instead of $\xi^{\prime}(a_0 =1)$. 
We note that the above choice was used recently by the authors of \cite{Wang:2018azy} and earlier in \cite{Li:2011ga,Guo:2017deu}.  

For the first interaction model (IVS1) it is possible to obtain an analytical solution for the background evolution of the dark sector fluids:
\begin{widetext}
\begin{eqnarray}
&&\rho_x = \rho_{x0} a^{3 (\xi_0 +\xi_a)} \exp\left(-3\xi_a (a-1)\right)~,\\
&&\rho_c = \rho_{c0} a^{-3} - 3 \rho_{x0} a^{-3} \int_{1}^{a}\left[\xi_0 + \xi_a (1-a)\right] a^{3 (\xi_0 +\xi_a)+2} \exp\left(-3\xi_a (a-1)\right)~.
\end{eqnarray}
\end{widetext}
For the IVS2 model the background evolution needs to be computed numerically.

To evaluate the perturbations equations in the presence of an interaction we work within the perturbed FLRW metric~\cite{Mukhanov,Ma:1995ey, Malik:2008im} and follow the synchronous gauge, see Ref.~\cite{Wang:2014xca} for details.  

Last but not least, we note that if in IVS1, the interaction function is simply considered as $Q = \Gamma \rho_x$, where $\Gamma$ is purely a time-independent constant, that means if the interaction rate depends solely on the dark energy component not on other physical parameters, such as the Hubble expansion rate,  scale factor of the universe, etc., it is then believed that such cosmological models should reflect more light on the intrinsic properties of the dark energy. Similarly,  without assuming any interaction in the dark sector, one could also construct some physically motivated phenomenological dark energy models where dark energy could decay with the evolution of the universe leading to a class of metabstable dark energy models \cite{Shafieloo:2016bpk,Landim:2016isc,Szydlowski:2017wlv,Li:2019san}.

\section{Observational data and statistical method}
\label{sec-data}

In this section we describe the cosmological observations that we have used to constrain the interacting scenarios. A discussion concerning the statistical method used in our analyses is also detailed.  
The publicly available datasets that we exploit in what follows are:

\begin{itemize}

\item \textbf{Cosmic Microwave Background (CMB)}: We use the Cosmic Microwave Background measurements from Planck 2015 data release~\cite{Adam:2015rua, Aghanim:2015xee},
 which include both the high-$\ell$ ($30\le \ell \le 2508$) TT and the low-$\ell$ ($2\le \ell \le 29$) TT likelihoods. The Planck polarization likelihood in the low-$\ell$ multipole regime ($2\le \ell \le 29$), together with the high-multipole ($30\le \ell \le 1996$) EE and TE likelihoods are also considered. Despite the fact that all these likelihoods have a clear dependence on a given number of nuisance parameters, such as residual foreground contamination, calibration, and others, we have also accounted for those in our numerical analyses and marginalized over them when presenting the final constraints.

\item \textbf{Baryon acoustic oscillation (BAO)} distance measurements:  we use the BAO data from different observational missions, see Refs.~\cite{Beutler:2011hx,Ross:2014qpa,Gil-Marin:2015nqa}. 

\item \textbf{Supernovae Type Ia (Pantheon)}:  The Supernovae Type Ia (SNIa) were the first indicators for an accelerating phase of the universe. Here, we use the latest compilation of SNIa data (known as Pantheon sample) comprising 1048 data points~\cite{Scolnic:2017caz}. 

\item \textbf{Hubble constant (R19)}: Finally, we shall also consider the impact of a recent estimation of the Hubble constant, $H_0 = 74.03 \pm 1.42$ km/s/Mpc at $68\%$ CL~\cite{Riess:2019cxk}, which shows a high tension ($4.4 \sigma$) with CMB estimates \textit{within the minimal $\Lambda$CDM cosmological model}.

We however combine these datasets in the context of our IVS1 and IVS2 dark matter-dark energy models, see Eqs.~(\ref{ivs1}) and (\ref{ivs2}), to explore whether this tension could be alleviated within these non-standard cosmologies.

\end{itemize}

For the statistical analyses, we make use of \texttt{cosmomc}, a Markov chain Monte Carlo package~\cite{Lewis:2002ah,Lewis:1999bs}, equipped with the Gelman and Rubin statistics for convergence diagnosis. This software also includes the support for the Planck 2015 likelihood~\cite{Aghanim:2015xee}. The parameter space we shall constrain is 
\begin{align}
\mathcal{P} \equiv\Bigl\{\Omega_{b}h^2, \Omega_{c}h^2, 100\theta_{MC}, \tau, \xi_0, \xi_{a}, n_{s}, log[10^{10}A_{S}]\Bigr\}~,
\label{eq:parameter_space}
\end{align}
in which  $\Omega_{b} h^2$ is the physical density for baryons; $\Omega_{c}h^2$ is the physical density for CDM; $\theta_{MC}$ denotes the ratio of sound horizon to the angular diameter distance; $\tau$ is the reionization optical depth; $n_{s}$ denotes the scalar spectral index; $A_S$ denotes the amplitude of the primordial scalar power spectrum and $\xi_0, \xi_{a}$ control the interaction rate among the two dark sectors. We are therefore exploring an eight-dimensional parameter space with two extra degrees of freedom compared to the six-dimensional $\Lambda$CDM cosmology. We note that in the present work we are considering the spatially flat FLRW universe.  Table \ref{tab:priors} presents the priors imposed on the model parameters for the statistical analyses.   

\begin{table}
\begin{center}
\begin{tabular}{c|c}
Parameter                    & Prior\\
\hline 
$\Omega_{b} h^2$             & $[0.005,0.1]$\\
$\Omega_{c} h^2$             & $[0.01,0.99]$\\
$\tau$                       & $[0.01,0.8]$\\
$n_s$                        & $[0.5, 1.5]$\\
$\log[10^{10}A_{S}]$         & $[2.4,4]$\\
$100\theta_{MC}$             & $[0.5,10]$\\ 
$\xi_0$                      & $[-1, 1]$\\
$\xi_a$                      & $[-1, 1]$\\ 
\end{tabular}
\end{center}
\caption{Flat priors imposed on various cosmological parameters of the interacting dark energy scenarios. }
\label{tab:priors}
\end{table}
\begingroup                                                                                                                     
\squeezetable                                                 
\begin{center}                              
\begin{table*}                                             
\begin{tabular}{cccccccccccccccc}       
\hline\hline                                                                                                                    
Parameters & CMB   & CMB+BAO & CMB+BAO+Pantheon & CMB+R19 & CMB+BAO+R19\\ \hline
$\Omega_c h^2$ & $    0.106_{ -0.040}^{+    0.039}$  & $    0.116_{-    0.044}^{  0.035}$ & $    0.121_{-    0.023}^{+    0.025}$ & $    0.084_{-    0.026}^{+    0.032} $& $    0.085_{-    0.030}^{+    0.037} $ \\

$\Omega_b h^2$ & $    0.02220_{-    0.00031}^{+    0.00032}$ &  $    0.02224_{-    0.00029}^{+    0.00031}$ & $    0.02226_{-    0.00031}^{+    0.00029}$ & $    0.02223_{-    0.00028}^{+    0.00031}$ & $    0.02223_{-    0.00030}^{+    0.00039}$\\

$100\theta_{MC}$ & $    1.0413_{-    0.0025}^{+    0.0027}$ &  $    1.0408_{-    0.0020}^{+    0.0027}$ & $    1.0404_{-    0.0015}^{+    0.0014}$ & $    1.0426_{-    0.0021}^{1+    0.0018}$ & $    1.0426_{-    0.0023}^{1+    0.0019}$\\

$\tau$ & $    0.079_{-    0.033}^{+    0.035}$ &  $    0.082_{-    0.036}^{+    0.035}$ &  $    0.083_{-    0.036}^{+    0.036}$ & $    0.077_{-    0.032}^{+    0.045}$ & $    0.078_{-    0.036}^{+    0.035}$\\

$n_s$ & $    0.9731_{-    0.0084}^{+    0.0088}$ & $    0.9747_{-    0.0084}^{+    0.0081}$ &   $    0.9750_{-    0.0084}^{+    0.0085}$ & $    0.9740_{-    0.0082}^{+    0.0086}$ &  $    0.9741_{-    0.0090}^{+    0.0090}$ \\

${\rm{ln}}(10^{10} A_S)$ & $    3.101_{-    0.065}^{+    0.066}$ &  $    3.106_{-    0.070}^{+    0.068}$ & $    3.108_{-    0.071}^{+    0.071}$ & $    3.096_{-    0.062}^{+    0.065}$ & $    3.100_{-    0.069}^{+    0.069}$\\

$\xi_0$ & $   -0.01_{-    0.21}^{+    0.21}$ &  $   -0.04_{-    0.20}^{+    0.23}$ &  $   -0.04_{-    0.17}^{+    0.15}$ & $   -0.03_{-    0.17}^{+    0.16}$ & $   0.05_{-    0.21}^{+    0.18}$ \\

$\xi_a$ & $    0.20_{-    0.56}^{+    0.71}$ &  $    0.18_{-    0.46}^{+    0.50}$ & $    0.14_{-    0.37}^{+    0.45}$ & $    0.55_{-    0.52}^{+    0.50}$ &  $    0.19_{-    0.48}^{+    0.52}$\\

$\Omega_{m0}$ & $    0.27_{-    0.12}^{+    0.13}$ &  $    0.30_{-    0.11}^{+    0.09}$ & $    0.313_{-    0.058}^{+    0.062}$ & $    0.202_{-    0.054}^{+    0.067}$ & $    0.214_{-    0.066}^{+    0.082}$\\

$\sigma_8$ & $    0.94_{-    0.28}^{+    0.35}$ &  $    0.87_{-    0.22}^{+    0.34}$ & $    0.82_{-    0.14}^{+    0.14}$ & $    1.09_{-    0.27}^{+    0.25}$ & $    1.10_{-    0.32}^{+    0.28}$\\

$H_0$ & $   69.2_{-    5.0}^{+    5.0}$&  $   68.5_{-    2.4}^{+    2.6}$ & $   68.0_{-    1.6}^{+    1.6}$ & $   72.9_{-    2.4}^{+    2.3}$ &. $   71.1_{-    1.9}^{+    2.3}$\\

$S_8$ & $    0.86_{-    0.09}^{+    0.13}$ &  $    0.84_{-    0.07}^{+    0.14}$ & $  0.827_{-    0.060}^{+    0.065}  $ & $    0.89_{-    0.11}^{+    0.10}$ & $    0.91_{-    0.11}^{+    0.09}$\\

\hline\hline                                                  
\end{tabular}                                                   
\caption{Mean values and 95$\%$~CL errors on the cosmological parameters for the IVS1 interacting scenario, $Q = 3 \left[\xi_0 +\xi_a (1-a)\right] H \rho_x$, using different combinations of the cosmological datasets considered here.}
\label{tab:Model1}                                              \end{table*}                                                    \end{center}                                                    \endgroup 
\begin{figure*}
\includegraphics[width=0.65\textwidth]{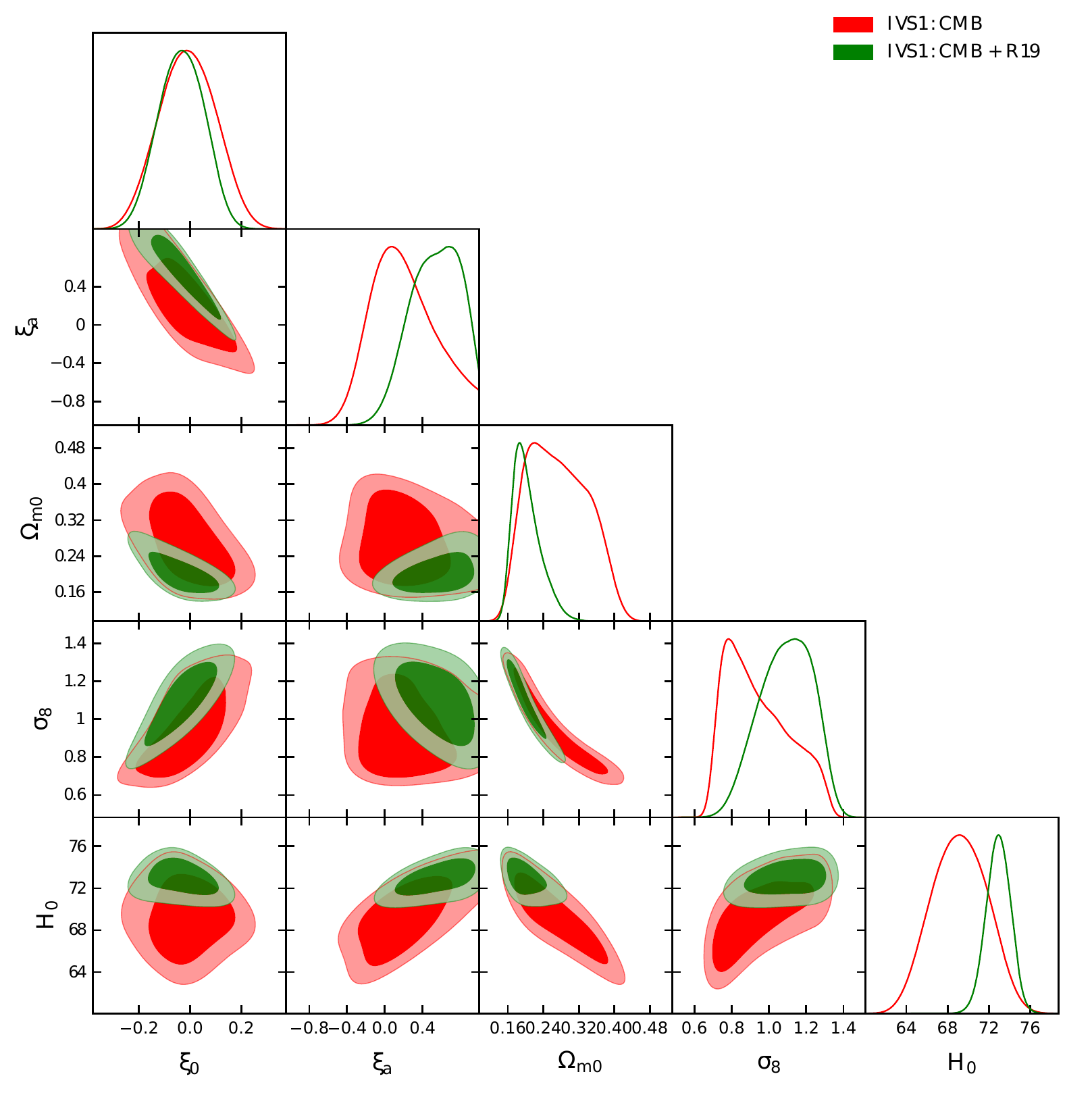}
\caption{Two-dimensional contours and one-dimensional marginalized posterior distributions for some key parameters of the IVS1 scenario for the CMB and CMB+R19 data sets.}
\label{Fig2D-ivs1C}
\end{figure*}
\begin{figure*}
\includegraphics[width=0.65\textwidth]{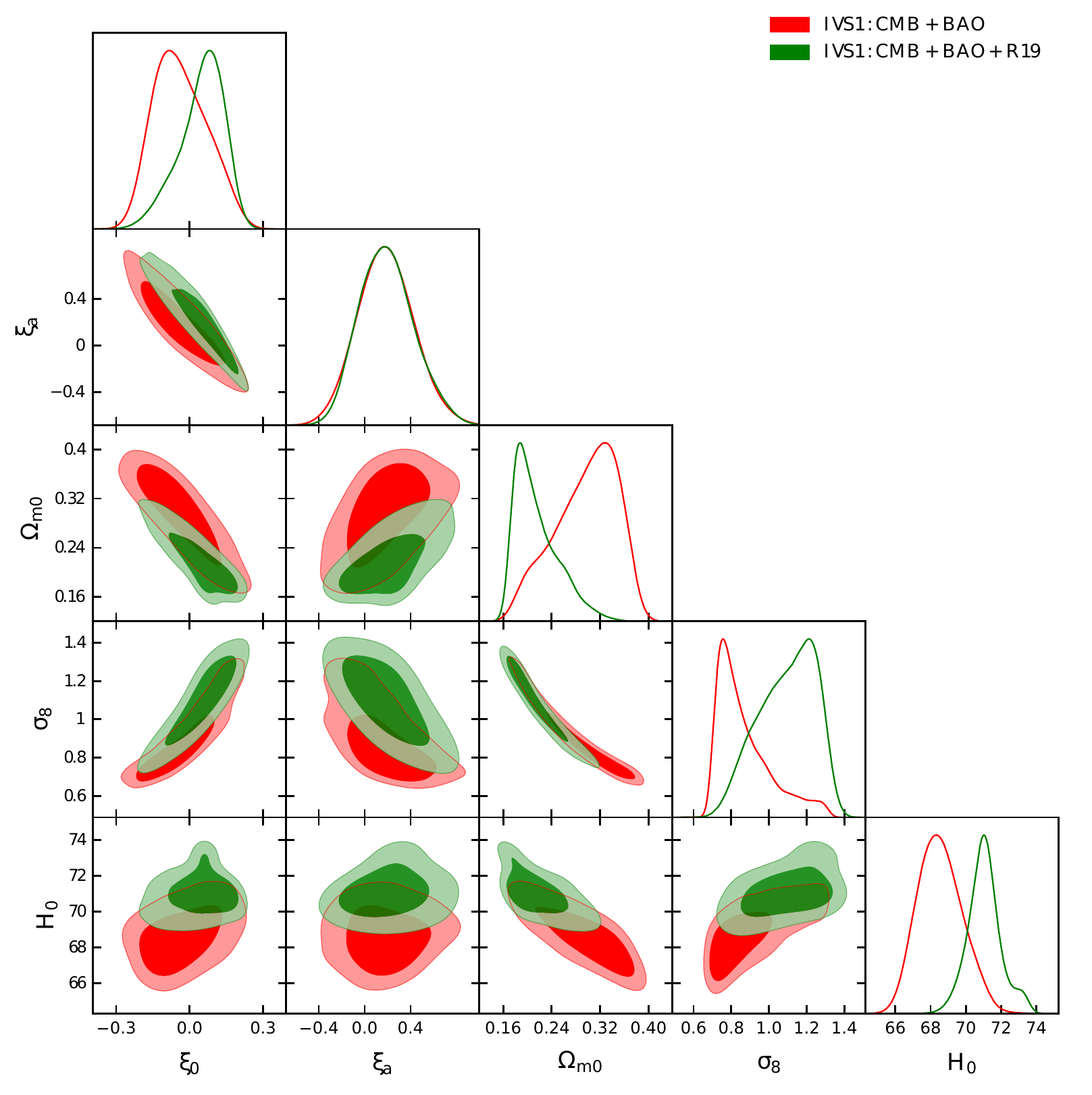}
\caption{Two-dimensional contours and one-dimensional marginalized posterior distributions for some key parameters of the IVS1 scenario for the CMB+BAO and CMB+BAO+R19 data sets.}
\label{Fig2D-ivs1CB}
\end{figure*}
\begin{figure*}
\includegraphics[width=0.4\textwidth]{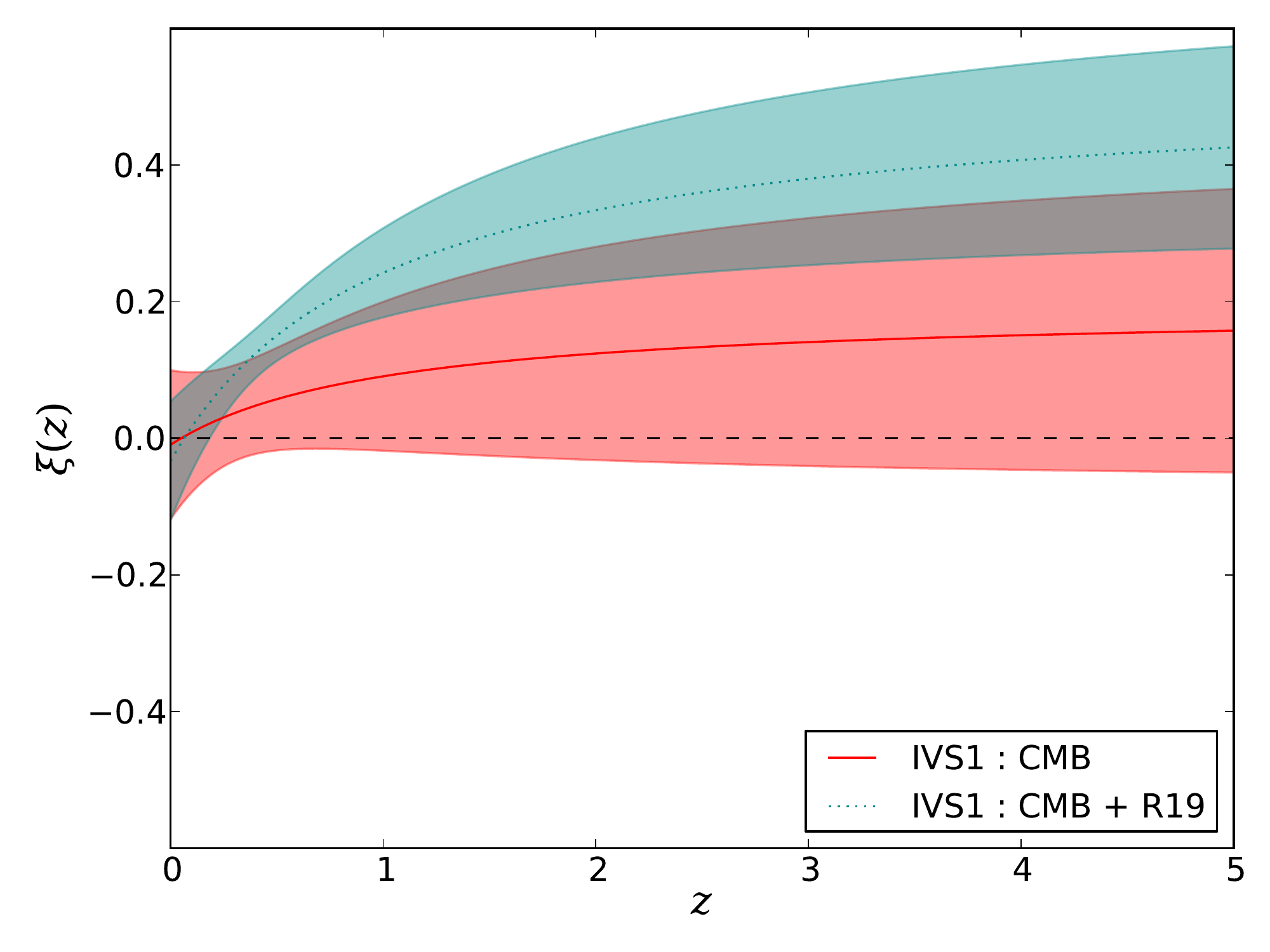}
\includegraphics[width=0.4\textwidth]{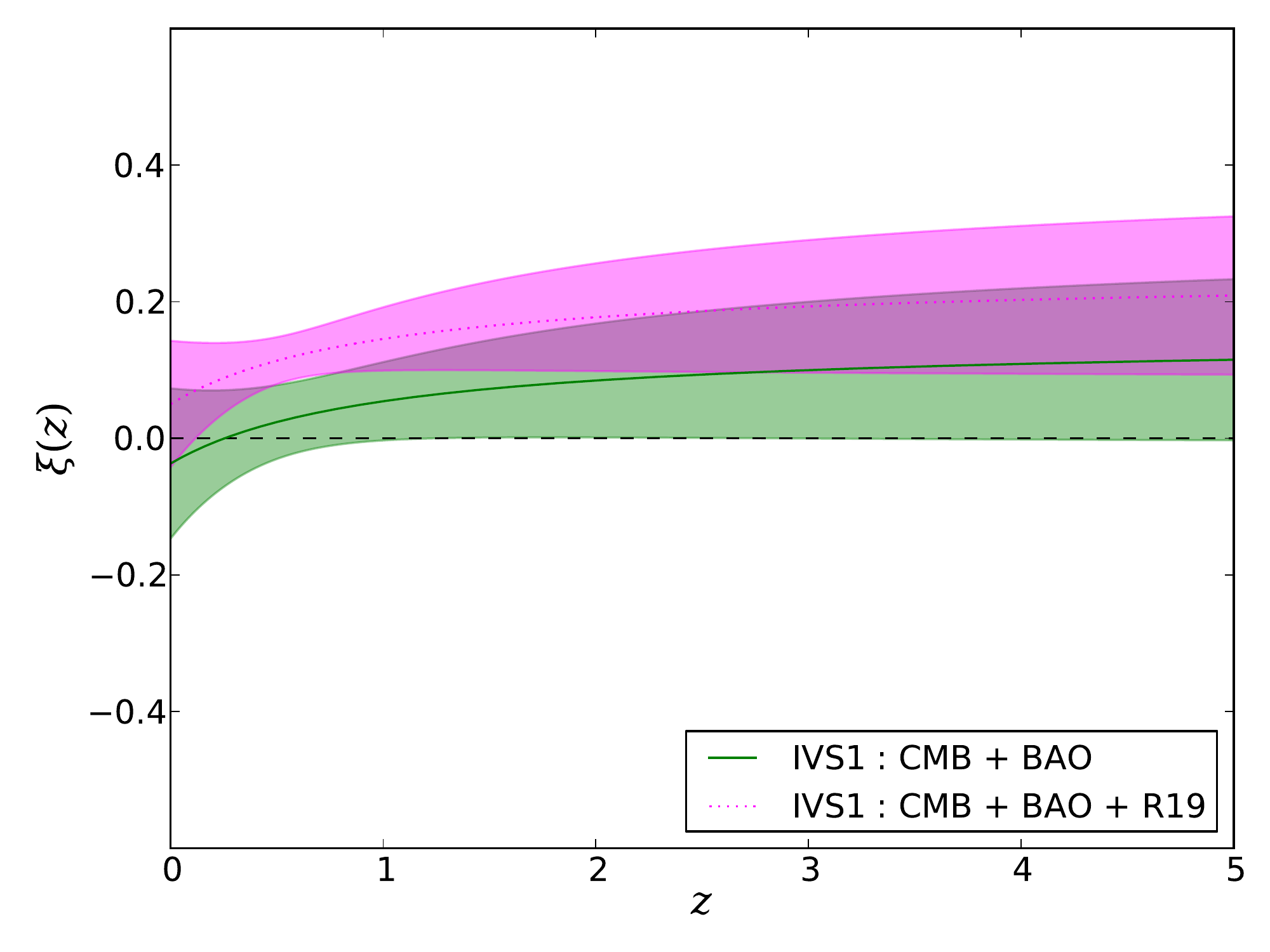}
\caption{Redshift evolution of the dynamical coupling parameter $\xi (z)$ ($1+z  = a^{-1}$) for the IVS1 scenario using various observational datasets. The left panel corresponds to the CMB and CMB+R19 datasets while the right panel stands for the CMB+BAO and CMB+BAO+R19. }
\label{xi_ivs1}
\end{figure*}

\section{Numerical analyses and results}
\label{sec-results}

 In this section we shall present the constraints on the interacting scenarios IVS1 and IVS2 (Eqs.~(\ref{ivs1}) and (\ref{ivs2}) respectively), arising from the combination of several datasets, namely, CMB,  CMB+R19, CMB+BAO, CMB+BAO+R19, and CMB+BAO+Pantheon.

\subsection{IVS1: $Q = 3 \left[ \xi_0 +\xi_a (1-a)\right] H \rho_x$}
\label{sec-model1}

The observational constraints for this interacting dark energy scenario have been displayed in Table~\ref{tab:Model1}. Figures~\ref{Fig2D-ivs1C} and~\ref{Fig2D-ivs1CB} depict the one-dimensional marginalized posterior distributions and the two-dimensional joint contours for some selected cosmological parameters. Notice from Tab.~\ref{tab:Model1} that the mean values of the parameters $\xi_0$ and $\xi_a$, quantifying the interaction among the dark sectors, are of opposite signs, and even if values $(\xi_0, \xi_a) \neq (0, 0)$ are still allowed by the observational data, a non-interacting scenario is consistent within 68\% CL. Figures~\ref{Fig2D-ivs1C} and \ref{Fig2D-ivs1CB} show the strong anti-correlation between the interaction parameters $\xi_0$ and $\xi_a$. 

Concerning the value of $H_0$ within the IVS1 interacting scheme, note that it is slightly larger than that obtained with Planck CMB data alone in the context of a $\Lambda$CDM model~\cite{Ade:2015xua}. Due to the larger error bars on $H_0$, the $4 \sigma$ tension between local measurements ($74.03\pm1.42$ km/s/Mpc~\cite{Riess:2019cxk}) and CMB observations is reduced to $2$ standard deviations. Combining the CMB dataset with a gaussian prior on $H_0$ from R19, see the fifth column of Tab.~\ref{tab:Model1}, we obtain $\xi_a = 0.55_{-0.52}^{+0.50}$ at $95\%$ CL, i.e. different from zero with a significance of 2 standard deviations. The dynamical evolution of the coupling parameter $\xi(z)$ is illustrated on the left panel of Fig.~\ref{xi_ivs1} considering the CMB (solid curve) and the CMB+R19 data combination (dotted curve), where all the parameters have been fixed to their mean values. As for the CMB+R19 data combination the dynamical coupling $\xi(z)$ is larger than zero, the matter density shifts towards a lower value (see the strong anti-correlation between $\Omega_{m0}$ and $\xi_0$ in Fig.~\ref{Fig2D-ivs1C}) and therefore there is not an increase in the quantity $S_8 \equiv\sigma_8 \sqrt{\Omega_{m0}/0.3}$. Consequently, in this case, the tension at more than $2\sigma$ on $S_8$ \cite{DiValentino:2018gcu} between Planck and the cosmic shear experiments, namely, KiDS-450~\cite{Kuijken:2015vca,Hildebrandt:2016iqg,Conti:2016gav}, DES~\cite{Abbott:2017wau,Troxel:2017xyo}, and CFHTLenS~\cite{Heymans:2012gg, Erben:2012zw,Joudaki:2016mvz} is solved within one standard deviation.

When adding the BAO dataset the error bars on the Hubble constant are notably decreased with respect to what we observed with the CMB alone \footnote{The inclusion of BAO data to other datasets has a notable effect on the error bars on $H_0$, see for instance \cite{Yang:2019bpr,Zhang:2018air,Yang:2018qmz}.}. In this case the tension with R19 is only mildly alleviated, as it remains present at the $3\sigma$ level. The addition of BAO measurements bring very close to zero the dynamical evolution of the dark sector coupling, see the right panel of Fig.~\ref{xi_ivs1}.

Finally, when adding the Pantheon dataset to CMB+BAO (i.e. the combination named as CMB+BAO+Pantheon in Tab.~\ref{tab:Model1}), we note that the estimation of $H_0$ shifts down and its error bars are reduced, increasing therefore the tension with R19 measurements.

\begingroup                                                                                                                     
\squeezetable                                                                                                                   
\begin{center}                                                                                                                  
\begin{table*}                                                                                                                   
\begin{tabular}{cccccccccccc}                                    
\hline\hline                                                    
Parameters & CMB  & CMB+BAO & CMB+BAO+Pantheon & CMB+R19&CMB+BA0+R19\\ \hline

$\Omega_c h^2$ & $    0.114_{-    0.048}^{+    0.043}$  & $    0.125_{-    0.036}^{+    0.031}$ & $    0.120_{-    0.032}^{+    0.031}$& $    0.081_{-    0.019}^{+    0.022}$ & $    0.093_{-    0.022}^{+    0.026}$\\

$\Omega_b h^2$ & $   0.02219_{-0.00030}^{+0.00031} $& $0.02225_{-0.00030}^{+0.00029} $ & $    0.02226_{-    0.00030}^{+    0.00030}$ & $    0.02223_{-    0.00028}^{+    0.00030}$ & $    0.02220_{-    0.00030}^{+    0.00028}$\\

$100\theta_{MC}$ & $    1.0408_{-    0.0025}^{+    0.0027}$ &  $    1.0402_{-    0.0018}^{+    0.0020}$ &  $    1.0405_{-    0.0016}^{+    0.0017}$ & $    1.0428_{-    0.0020}^{+    0.0013}$ & $    1.0420_{-    0.0016}^{+    0.0015}$\\

$\tau$ & $    0.080_{-    0.034}^{+    0.034}$ &  $    0.083_{-    0.034}^{+    0.034}$ & $    0.084_{-    0.034}^{+    0.033}$ & $    0.086_{-    0.032}^{+    0.030}$ & $    0.080_{-    0.037}^{+    0.037}$\\

$n_s$ & $    0.9733_{-    0.0085}^{+    0.0095}$ &  $    0.9749_{-    0.0088}^{+    0.0087}$ & $    0.9755_{-    0.0078}^{+    0.0081}$ & $    0.9751_{-    0.0082}^{+    0.0081}$ & $    0.972_{-    0.010}^{+    0.011}$\\

${\rm{ln}}(10^{10} A_S)$ & $    3.105_{-    0.067}^{+    0.066}$ &  $    3.108_{-    0.066}^{+    0.066}$ & $    3.109_{-    0.066}^{+    0.064}$ &  $    3.114_{-    0.067}^{+    0.066}$ & $    3.106_{-    0.072}^{+    0.073}$ \\

$\xi_0$ & $    -0.02_{-    0.38}^{+    0.43}$ &  $   -0.17_{-    0.38}^{+    0.41}$ & $   -0.06_{-    0.40}^{+    0.36}$ &  $    0.15_{-    0.30}^{+    0.32}$ & $    0.07_{-    0.37}^{+    0.40}$\\

$\xi_a$ & $    0.22_{-    0.59}^{+    0.78}$ &  $    0.41_{-    0.59}^{+    0.59}$ & $    0.17_{-    0.45}^{+    0.83}$ &  $    0.39_{-    0.74}^{+    0.69}$ & $    0.37_{-    0.87}^{+    0.81}$\\

$\Omega_{m0}$ & $    0.30_{-    0.15}^{+    0.15}$ &  $    0.321_{-    0.098}^{+       0.094}$ & $    0.312_{-    0.097}^{+    0.077}$ & $    0.195_{-    0.044}^{+       0.053}$ & $    0.229_{-    0.052}^{+       0.062}$\\

$\sigma_8$ & $    0.86_{-    0.17}^{+    0.19}$ & $    0.82_{-    0.11}^{+      0.12}$ & $    0.83_{-    0.10}^{+    0.12}$ &  $    0.997_{-    0.088}^{+      0.083}$ & $    0.95_{-    0.10}^{+      0.09}$ \\

$H_0$ & $   68.3_{-  6.2}^{+    6.0}$  & $   68.0_{-    2.6}^{+    2.9}$ & $   67.9_{-    2.2}^{+    2.6}$ & $   73.1_{-  2.4}^{+    2.3}$ & $   71.3_{-  2.2}^{+    2.2}$ \\

$S_8$ & $    0.846_{-    0.068}^{+     0.046}$ &  $    0.845_{-    0.039}^{+     0.035}$ & $  0.840_{-    0.039}^{+    0.034}  $ & $    0.800_{-    0.047}^{+    0.050}$ & $    0.826_{-    0.042}^{+    0.041}$\\
\hline\hline  

\end{tabular}                                                                                                             
\caption{Mean values and 95$\%$~CL errors on the cosmological parameters of the IVS2 interacting scenario, $Q = 3 \left[\xi_0 +\xi_a (1-a)\right] H \frac{\rho_c\rho_x}{\rho_c+\rho_x}$, using different combinations of the cosmological datasets considered here.}
\label{tab:Model2}                                                                                                   
\end{table*}                                                                                                                     
\end{center}                                                                                                                    
\endgroup 
\begin{figure*}
\includegraphics[width=0.65\textwidth]{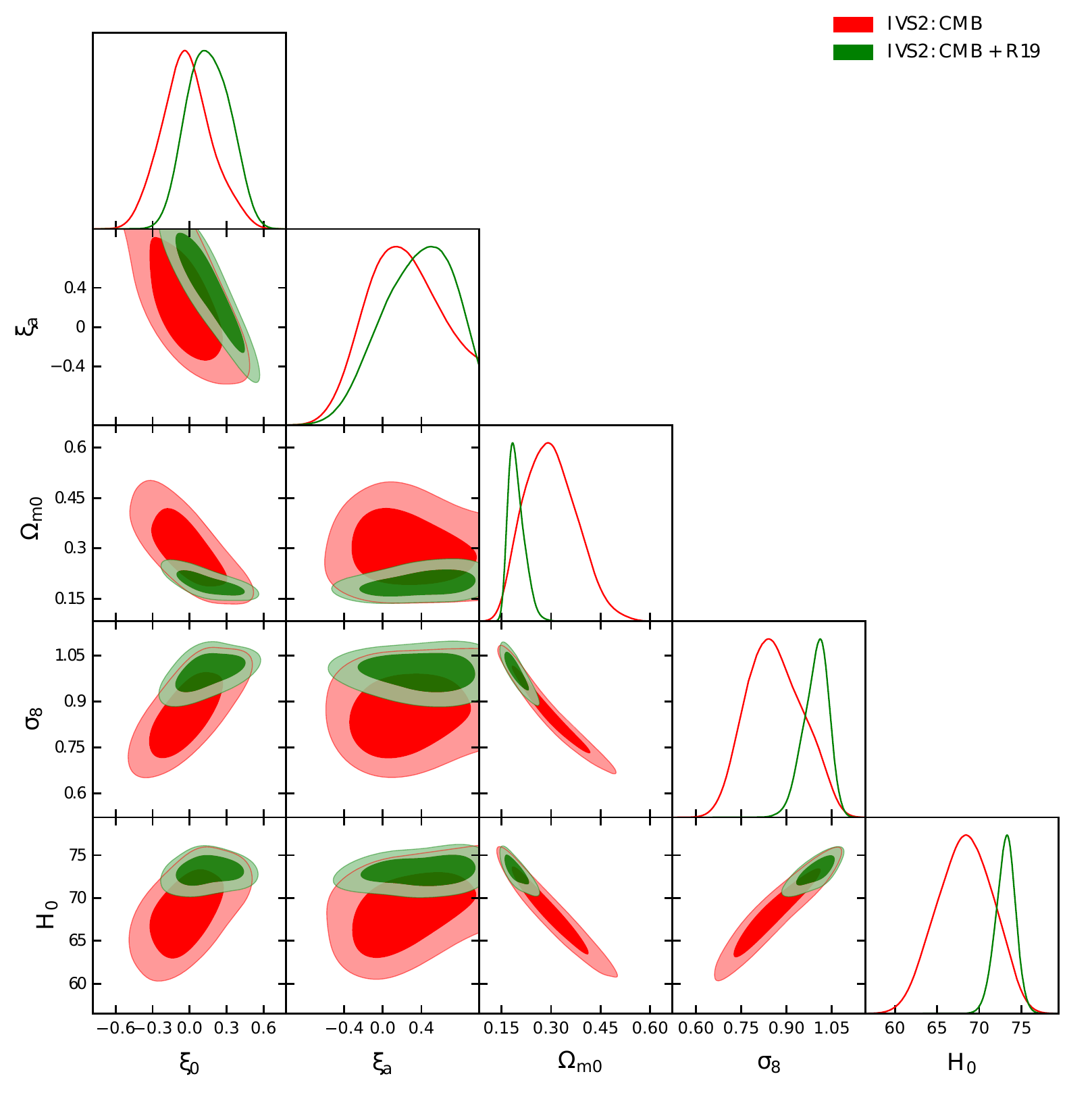}
\caption{Two-dimensional contours and one-dimensional marginalized posterior distributions for some key parameters of the IVS2 scenario for  the CMB and CMB+R19 data sets.}

\label{Fig2D-ivs2C}
\end{figure*}
\begin{figure*}
\includegraphics[width=0.65\textwidth]{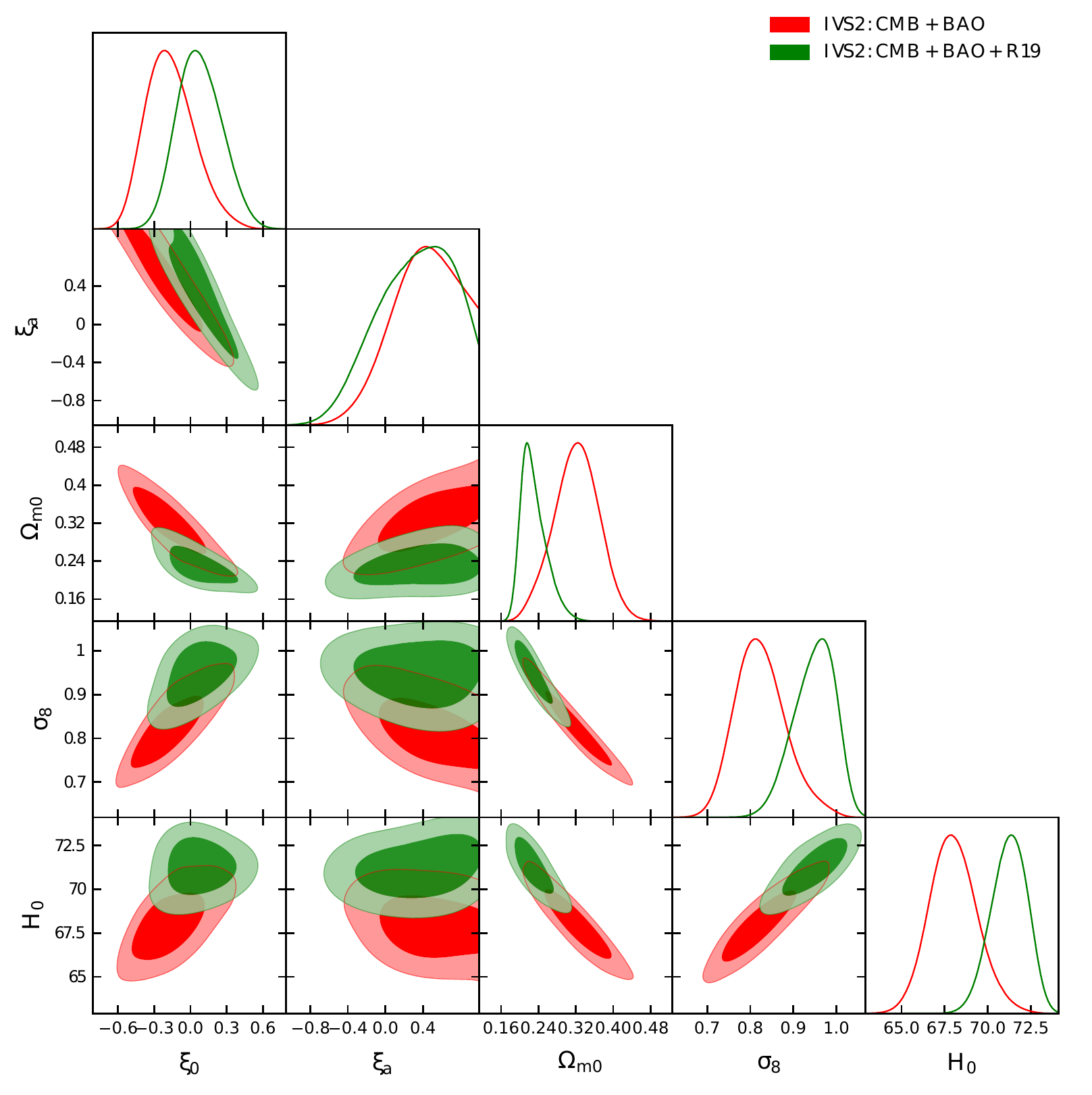}
\caption{Two-dimensional contours and one-dimensional marginalized posterior distributions for some key parameters of the IVS2 scenario for the CMB+BAO and CMB+BAO+R19 data sets.}

\label{Fig2D-ivs2CB}
\end{figure*}
\begin{figure*}
\includegraphics[width=0.4\textwidth]{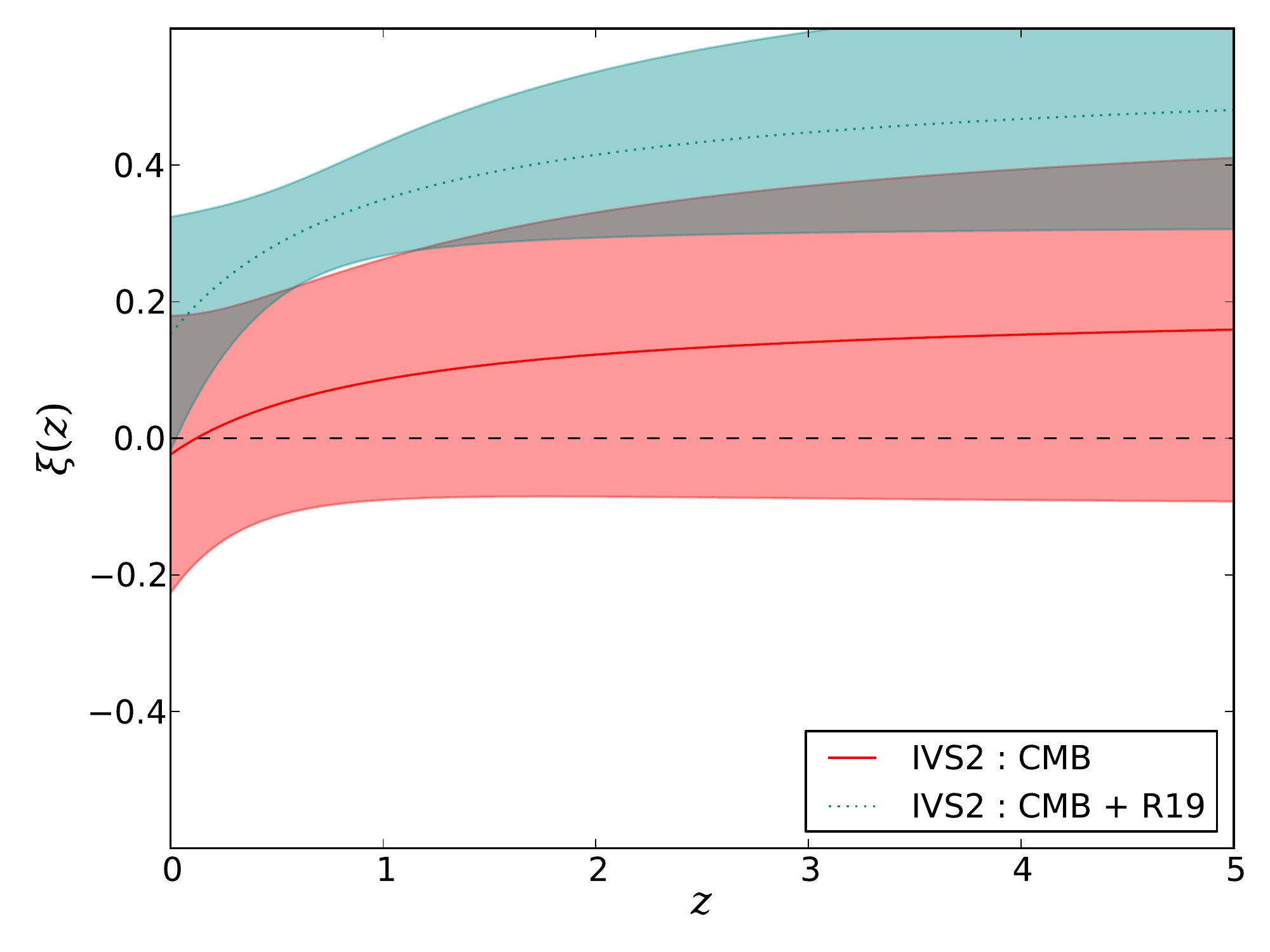}
\includegraphics[width=0.4\textwidth]{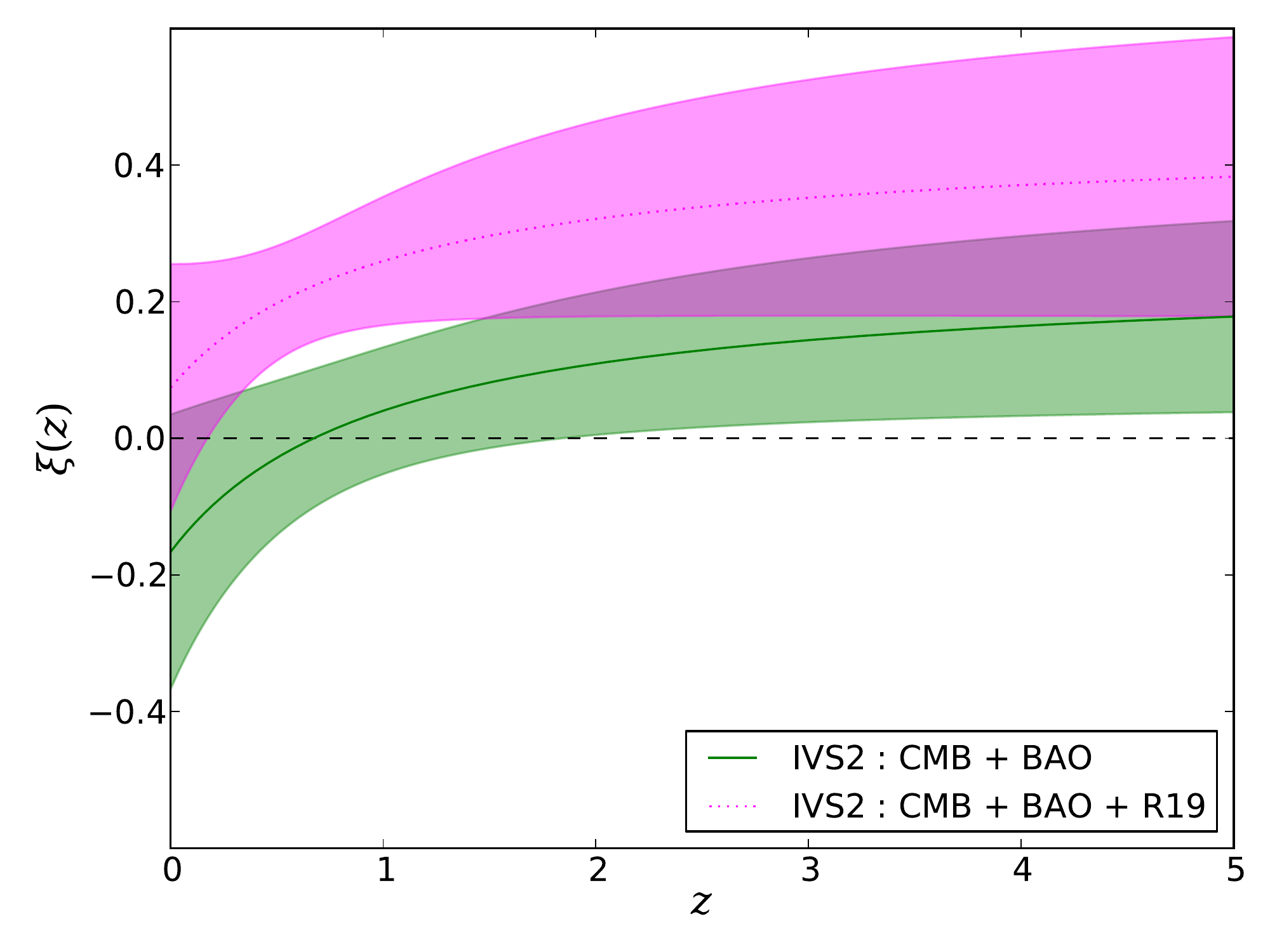}
\caption{We show the qualitative evolution of the dynamical coupling parameter $\xi (z)$ ($1+z  = a^{-1}$) for the IVS2 scenario using various observational datasets. The left panel corresponds to the CMB and CMB+R19 datasets while the right panel stands for the CMB+BAO and CMB+BAO+R19. }
\label{xi_ivs2}
\end{figure*}

\subsection{IVS2: $Q = 3 \left[ \xi_0 +\xi_a (1-a)\right] H \frac{\rho_c\rho_x}{\rho_c+\rho_x}$}
\label{sec-model2}

The summary of the observational constraints on this interaction scenario is shown in Tab.~\ref{tab:Model2}, while in Figs.~\ref{Fig2D-ivs2C} and \ref{Fig2D-ivs2CB} we depict the one-dimensional marginalized posterior distributions and the two-dimensional contour plots for a number of both independent and derived cosmological parameters, emphasizing their correlations with $\xi_0$ and $\xi_a$.
Note that for CMB alone data the mean value of $\xi_0$ is almost zero with a very mild preference for negative values, while the mean value of $\xi_a$ is found to be positive. When external datasets such as BAO or BAO plus Pantheon are added to CMB observations the tendency of  $\xi_0$ ($\xi_a$) to take negative (positive) values is enhanced, showing opposite behaviors and a strong negative correlation between them, as we can see from Figs.~\ref{Fig2D-ivs2C} and \ref{Fig2D-ivs2CB} regardless of the observations considered in the analysis. Furthermore, from Tab.~\ref{tab:Model2}, it is also possible to notice that for all the observational datasets the values $(\xi_0, \xi_a)  = (0, 0)$ are allowed within 95\% CL implying that we recover the non-interacting $\Lambda$CDM limit. Nevertheless the dynamical interacting scenario cannot be ruled out and indeed, based on present observations, no definite conclusion can be made. 

Focusing on the estimation of $H_0$ for this interaction scenario with CMB data only (see Tab.~\ref{tab:Model2}), we notice that it is slightly larger than within the $\Lambda$CDM scenario, and its error bars are increased due to the presence of a dynamical coupling. Due to the larger error bars on $H_0$, for this model it is also possible to solve the tension with the local measurements of R19~\cite{Riess:2019cxk}. Combining the CMB dataset with a gaussian prior on $H_0$ from R19 provides therefore a possible solution of both the Hubble constant tension and the $S_8$ tension with the cosmic shear data. The dynamical evolution of the coupling parameter $\xi(z)$ for this combination of data is shown in Fig.~\ref{xi_ivs2}, notice that in this case there is also a tendency for $\xi(z) >0$ for CMB + R19 and the addition of BAO data is less able to restore $\xi(z) =0$ than within the IVS1 scenario. For CMB+BAO or CMB+BAO+Pantheon the situation is very similar to the dynamical IVS1 scenario aforementioned, i.e. neither the $H_0$ nor the $S_8$ tension are alleviated.

Finally, we comment on the results from a Bayesian evidence analysis of the dynamical interacting scenarios here explored. In this framework, a comparison of a cosmological model is performed with respect to a standard and well-motivated cosmological model~\cite{Heavens:2017afc,Heavens:2017hkr,Pan:2017zoh,Yang:2018qmz}. The $\Lambda$CDM provides the ideal choice for such a comparison.  We can introduce the so-called Jeffreys scale, which, for different possible values of $\ln B_{ij}$, quantifies the strength of evidence of the reference, the canonical $\Lambda$CDM scenario ($M_i$), with respect to the underlying cosmological model ($M_j$) \cite{Kass:1995loi}.
We have for $0 \leq \ln B_{ij} < 1$, a weak evidence; for $1 \leq \ln B_{ij} < 3$, a Definite/Positive evidence; for $3 \leq \ln B_{ij} < 5$, a strong evidence; and for $\ln B_{ij} \geq 5$, a very strong evidence for the $\Lambda$CDM model against the underlying cosmological scenario, here the interacting scenario. 
Following \cite{Heavens:2017afc,Heavens:2017hkr} we compute the values of $\ln B_{ij}$ for all the observational datasets employed in this work, and present the results in Tab.~\ref{tab:bayesian}. From this table, we learn that the $\Lambda$CDM model is always preferred over the two IVS models analyzed here. This is not surprising because $\Lambda$CDM has six free parameters while the two IVS models (IVS1 and IVS2) have eight free parameters. This fact eventually favors the base $\Lambda$CDM cosmology over the IVS models.

\begin{table}      
\begin{center}                    
\begin{tabular}{ccccccccc}                                      \hline\hline               
Dataset & Model &~~ $\ln B_{ij}$ & \\ 
\hline
CMB & IVS1 & $5.9$ \\
CMB+BAO & IVS1 & $8.5$ \\
CMB+BAO+Pantheon & IVS1 & $4.1$  \\
CMB+R19          & IVS1 & $2.8$\\
CMB+BAO+R19      & IVS1 & $9.9$\\
\hline\hline 
CMB & IVS2 & $5.1$  \\
CMB+BAO & IVS2 & $7.7$  \\
CMB+BAO+Pantheon & IVS2 & $3.8$  \\
CMB+R19          & IVS2 & $3.3$\\
CMB+BAO+R19      & IVS2 & $10.1$\\
\hline \hline 

\end{tabular}    
\caption{The table summarizes the values of $\ln B_{ij}$ computed for the $\Lambda$CDM model with respect to the IVS models.} 
\label{tab:bayesian}                          
\end{center}    
\end{table}

\section{Summary and Conclusions}
\label{sec-conclu}

Non canonical cosmologies with an interaction between the dark matter and dark energy fluids have been widely investigated in the past several years. From the observational perspective, interacting theories have been found to provide a very promising way to solve the tension between early and late universe cosmological estimates of the Hubble constant.  The present work generalizes interacting dark matter-dark energy models by considering a dynamical, redshift-dependent, coupling parameter. 

In mostly all of the dark sector interacting theories, characterized by exchange rates $Q = 3 H \xi f (\rho_c, \rho_x)$ (where $f$ is an analytic function of the arguments $\rho_c$ and/or $\rho_x$), the coupling parameter $\xi$ is assumed to be independent of time (see however e.g. Ref.~\cite{Wang:2018azy}). Unless one is interested in minimizing the number of extra parameters in the theory, there exists no underlying  symmetry or law in nature  which forbids such a dynamical coupling parameter. We have considered a very natural functional form for $\xi  = \xi_0 + \xi_a (1-a)$, that we have embedded into two possible interaction models, IVS1 ($Q = 3 \left[ \xi_0 +\xi_a (1-a)\right] H \rho_x$) and  IVS2 ($Q = 3 \left[ \xi_0 +\xi_a (1-a)\right] H \frac{\rho_c\rho_x}{\rho_c+\rho_x}$).

We find that the interaction parameters $\xi_0$ and $\xi_a$, governing the dynamical behavior of the coupling $\xi$, are, in almost all cases, perfectly compatible with a non interacting scenario, showing a strong negative correlation among them. Nevertheless, for the CMB+R19 data combination, we find an indication for $\xi_a>0$ at more than $2\sigma$ CL for IVS1. More importantly, when considering CMB data alone, we find, in general, that $\xi(z)>0$, leading to a smaller value of the present matter density. In order to leave the CMB acoustic peaks location unchanged (which are mostly sensitive to the $\Omega_m h^2$ combination), a larger value of the Hubble constant $H_0$ is required. This in turn implies an optimal scenario where to address both the $H_0$ and $S_8$ tensions between early and late universe's observations.
 
 Even if a Bayesian evidence analysis taking into account all observational datasets shows no particular preference for these interacting dark matter-dark energy models, a dynamical character in the interaction functions is still allowed by observations and can solve some pending issues related to high and low-redshift cosmological tensions.

\acknowledgments 
The authors thank the referees for their important comments and suggestions aiming to improve the manuscript. WY acknowledges the support from the National Natural Science Foundation of China under Grants No.  
11705079 and No.  11647153. OM is supported by the Spanish grants FPA2017-85985-P and SEV-2014-0398 of the MINECO and the European Union's Horizon 2020 research and innovation program under the grant agreements No.\ 690575 and 674896. 
SP gratefully acknowledges several fruitful discussions with Prof. J. D. Barrow on this topic. The research of SP has been supported by the Mathematical Research Impact-Centric Support Scheme (MATRICS), File No. MTR/2018/000940, given by the Science and Engineering Research Board (SERB), Govt. of India, and also by the Faculty Research and Professional Development Fund (FRPDF) Scheme of Presidency University, Kolkata, India.  
EDV acknowledges support from the European Research Council in the form of a Consolidator Grant with number 681431.

\end{document}